\documentclass[reviewcopy]{elsart}

\usepackage{psfig}

\begin{document}

\begin{frontmatter}

\title{Epitope analysis for influenza vaccine design}

\author{
Enrique T.\ Mu\~noz and Michael W.\ Deem}

\address{
Department of Bioengineering and
Department of Physics \& Astronomy\\
Rice University\\
Houston, TX 77005--1892
}
\bigskip

\address{
Tel: 713--348--5852\\
FAX: 713--348-5811\\
mwdeem@rice.edu
}

\newpage

\begin{abstract}
Until now, design of the annual influenza vaccine has
relied on phylogenetic or whole-sequence comparisons 
of the viral coat proteins hemagglutinin and neuraminidase,
with vaccine effectiveness assumed to correlate monotonically
to the vaccine-influenza sequence difference.
We use a theory from statistical mechanics to quantify
the non-monotonic immune response that results from antigenic
drift in the epitopes of the hemagglutinin and neuraminidase
proteins.
The results 
explain the ineffectiveness of the 2003--2004 influenza vaccine
in the United States
and
provide an accurate measure by which to optimize
the effectiveness of future annual influenza vaccines.
\end{abstract}

\end{frontmatter}

\newpage

\section{Introduction}

Antigenic variation constitutes one mechanism employed by influenza viruses
to evade the adaptive response of the host immune system.
This antigenic drift of the recognized, epitope regions
of the viral surface proteins hemagglutinin (HA) and neuraminidase (NA)
constitutes a major challenge to effective vaccine design, 
where historical experience and phylogenetic
analysis of HA and NA protein sequences from circulating 
human strains are used to decide the components of
the annual influenza vaccine.  Here we introduce a theory
to guide this important public health decision.
Application of the theory could
help prevent critical situations such as
occurred with 2003--2004 influenza epidemic \cite{CDC}, whence the 
administered A/Panama/2007/99 
H3N2 vaccine gave unexpectedly \cite{WHO}
low protection against the
mutant strain A/Fujian/411/2002.
A model from statistical mechanics
is used to evaluate the 
non-linear decrease of the immune response due to
mutations in the viral epitope region sequences.
We propose that this epitope analysis be regularly used 
as a measure of the immunological distance between mutant strains
in the annual design of the influenza vaccine.

Influenza A virus infections and posterior complications, such as
pneumonia, are a major cause of human morbidity and mortality. The 
2003--2004 influenza epidemic
was mainly due to the proliferation of the new H3N2 subtype
strain A/Fujian/411/2002, an antigenic drift mutant of A/Panama/2007/99. 
According to the February 2003 WHO report \cite{WHO}, after comparing the whole 
hemagglutinin (HA) sequences of both strains, the CDC council members 
concluded that both proteins were similar enough to expect a significant
degree of 
cross protection, and decided to include the
Panama strain in the H3N2 component of the vaccine.
No information 
concerning the neuraminidase sequence for the
Fujian strain was used. Recent clinical results, as stated in the 
16 January 2004 CDC Morbidity and Mortality 
Report \cite{CDC}, show the vaccine provided essentially no  
protection against infection during the 2003--2004 season.

\section{Methods}

We have developed
a theory of the immune response to an antigenic drift strain after vaccination 
based on statistical mechanics, Figure \ref{fig1} \cite{Deem}.  The model 
predicts the affinity constant values 
\begin{equation}
 K^{{\rm eq}} = \frac{\left[ {\rm Antigen}:{\rm Antibody} \right]}{\left[ {\rm Antigen} \right] \left[ {\rm Antibody} \right]}
\end{equation}
for a second antigen, after exposure to an original antigen whose epitope 
region differs by probability $p_{\rm epitope}$.
The key measure of antigenic drift in the theory is 
$p_{\rm epitope}$,
 the fractional change between the dominant epitope regions of 
the vaccine and the circulating strain, defined by the equation
\begin{equation}
 p_{\rm epitope} =
    \frac{{\rm number\: of\: mutations\: within\: the\: epitope}}
   {{\rm number\: of\: amino\: acids\: within\: the\: epitope}} \ .
\end{equation}
This characterization of antigenic drift in our 
theoretical model emphasizes the experimental fact that only the
epitope regions 
are significantly involved in immune recognition, as shown by 
immunoassays and crystallographic images \cite{Epitope2}. 

In the theory, it is the percent of the epitope that changes that
characterizes antigenic drift.
To provide additional empirical support for the theory,
we performed 
an historical analysis of the influenza seasons between 
1991--2000 when the H3N2 virus suptype was dominant
\cite{MMWR91,MMWR92,MMWR93,MMWR94,MMWR95,MMWR96,MMWR97,MMWR98,MMWR99}.
For every season, 
hemagglutinin sequences \cite{Lanlha2} were compared between the vaccine 
strain and the predominant circulating strain. 
A quantitative scale was defined 
to measure the seasonal flu severity as follows: low (1), mild (2) and high (3).
The values of seasonal flu severity were
correlated with the calculated  $p_{\rm epitope}$
 values.  In addition, to make clear that the epitope region
is primarily responsible for immune recognition, we also correlated
the seasonal flu severity with antigenic drift of the
entire hemagglutinin sequence,
normalized by the total number of amino acids in the protein
$p_{{\rm sequence}}$.
The results presented in Figure \ref{fig2} show that
seasonal flu severity is correlated with
$p_{{\rm epitope}}$ rather than $p_{{\rm sequence}}$,
thus favoring the epitope analysis approach.
Therefore, it is both logical and consistent with the observed data
to characterize antigenic drift by
the number of mutations within the epitope regions,
as we do in the present work. 
 
According to historic clinical 
experience, and to our model, when the antigenic drift between the vaccine 
and circulating strain, characterized by the $p_{\rm epitope}$ value, 
is small, exposure to the vaccine antigen
leads to a higher affinity constant than
without exposure.  This result is why immune system memory and vaccination are
generally effective. When the antigenic drift
between the vaccine and circulating strain is large, the vaccine antigen
is uncorrelated with the circulating strain antigen, and so immune 
system memory does not play a role. When the antigenic drift
between the vaccine and circulating
strain epitopes is 
modest $(0.23 < p_{\rm epitope} < 0.6)$, our
theory predicts that memory response may be worse than the naive response
(the solid curve lies below 
the dashed curve in Figure \ref{fig1}), which means that the immunological 
memory from the vaccine exposure actually gives worse protection, i.e., a lower 
affinity constant, than would no vaccination whatsoever.
This result is the original antigenic sin phenomena for influenza: 
vaccination creates memory sequences that for some mutation rates of influenza 
may increase susceptibility to future exposure \cite{Fazekas1,Davenport}. 
Parenthetically, not every infectious disease exhibits original antigenic sin, 
with measles one such example. The measles virus does not undergo 
antigenic drift, and despite approximately eight different subtypes 
that have been identified to date \cite{Bellini}, the HA and NA genetic
variation among 
them does not exceed 7\% on a nucleotide basis \cite{Bellini}, or roughly
2\% on an amino acid basis. Accordingly, within the context of our model, 
there is no possibility of original antigenic sin for measles
(since $p_{\rm epitope} < 0.02$, see Figure \ref{fig1}).
   
Original antigenic sin stems from localization of the immune system 
response in antibody sequence space. This localization is a result of the 
roughness in sequence space of the evolved antibody affinity constant for
antigen.  
Interestingly, there appears to have been a modest degree of original
antigenic sin, termed negative vaccine effectiveness in the CDC Morbidity
and Mortality report \cite{CDC},
associated with the 2003--2004 influenza vaccine.

Human influenza A viruses are classified in different subtypes according to 
the neuraminidase and hemagglutinin proteins. The current influenza A vaccine
includes both the H1N1 and H3N2 subtypes, and the
consensus sequence for the HA and NA proteins corresponding to each subtype 
requires annual update due to continuous antigenic drift. 
Variations due to point mutations 
in the residues in the epitope regions can considerably reduce 
the immune response, despite
biochemical cross activity between strains 
related by antigenic drift.
The epitope regions of the HA and NA proteins are shown in
Figure \ref{fig3}.
We propose that antigenic drift mutants be compared 
not by the whole sequences of the HA and NA proteins, but more precisely 
by the sequences of the dominant epitopes for the proposed vaccine strain, by
calculating the $p_{\rm epitope}$ parameter of our theory.
According to the definition (2), a different value of $p_{\rm epitope}$ is
obtained for each epitope region in both hemagglutinin and neuraminidase 
viral proteins. We propose to include in the analysis
 only the $p_{\rm epitope}$ values 
corresponding to the dominant epitopes in both proteins.  Since
both hemagglutinin and neuraminidase participate in the immune 
recognition process, it is some combination of the immune recognition
of these two proteins that contributes to reducing the seasonal
flu severity.  We, thus, define an approximate total response
as the binding constant at the average of the 
$p_{\rm epitope}$ values for the dominant HA and NA epitopes:
$p_{\rm avg} =
   \frac{1}{2} (p_{\rm epitope}^{\rm HA} + p_{\rm epitope}^{\rm NA} )$.

\section{Results}

We compared the epitope sequences of the HA protein to look for mutations in
the A/Fujian/411/2002 strain \cite{Lanlha2} with
respect to the
A/Panama/2007/99 strain \cite{Lanlha1}. The hemagglutinin H3 protein 
has five epitope regions (A, B, C, D, E) that have been identified and 
sequenced \cite{Lanlha}, among
which A and B are usually dominant \cite{Bush99,Cox}.
There exists experimental and clinical evidence that
epitope regions mutate much faster than other regions in the viral proteins,
presumably due to antibody selective pressure \cite{Fitch,Bush,Plotkin},
with the dominant epitopes mutating most rapidly \cite{Fitch2}.
Therefore, in the absence of more detailed information, 
we take an observed high
mutation rate (i.e., a high $p_{\rm epitope}$ value) in a given 
epitope to correlate with dominance.
Epitope A (residues 122, 124, 126, 130--133, 135, 137, 138, 140,
142--146, 150, 152, 168) presents one point mutation at residue 131. The 
calculated value $p_{\rm epitope} = 1/19 = 0.053$. Epitope B 
(residues 128, 129, 155--160, 163, 165, 186--190, 192--194, 196--198) presents 
three point mutations at residues 155, 156, and 186. The calculated value
$p_{\rm epitope} = 3/21 = 0.14$. Epitope C 
(residues 44--48, 50, 51, 53, 54, 273, 275, 276, 278--280, 294, 297, 299, 300,
304, 305, 307--312) presents one point
mutation at residue 50. 
The calculated value $p_{\rm epitope} = 1/27 = 0.037$. Epitope D 
(residues 96, 102, 103, 117,
121, 167, 170--177, 179, 182, 201, 203, 207--209, 212--219, 226--230,
238, 240, 242, 244, 246--248) presents no mutations. Epitope E
(residues 57, 59, 62, 63, 67, 75, 78, 80--83, 86--88, 91, 92, 94, 109,
260--262, 265) presents two point mutations at residues 75 
and 83. The calculated value $p_{\rm epitope} = 2/22 = 0.09$.
In absence of further information, which is the typical case
for the annual task of influenza vaccine design, we conclude that
likely B epitope is dominant 
and E epitope is subdominant for the A/Panama/2007/99
HA protein, whereas the other epitopes are cryptic. 
The dominance of epitope B is
in accordance with observed data \cite{Kostolansky}.
Note that by looking at the antigenic drift within the dominant
epitope, rather than the drift of the whole protein sequence,
we obtain a larger and much more accurate estimate of
the degree to which the immune response to 
A/Fujian/411/2002 and A/Panama/2007/99 will differ.

We also calculated the values for antigenic drift in the NA epitopes
between the A/Fujian/411/2002 \cite{Lanlna2} and 
A/Panama/2007/99 \cite{Lanlna1} strains.
The neuraminidase N2 protein has been completely sequenced and 
crystallized \cite{Tulip,Air1995}.
Mutational studies with
monoclonal antibodies identified three 
regions (A,B,C) in NA N2 that are important for recognition 
\cite{Epitope2,Gulati}, of which only the surface residues can be within
the epitopes. Regions  A and B are usually dominant \cite{Gulati}. 
Epitope A (residues 383--387, 389--394, 396, 399, 400, 401, 403)
presents three point mutations in at residues 385, 399, and
403. The calculated value $p_{\rm epitope} = 3/16 = 0.188$. Epitope B
(residues 197--200, 221, 222) presents
two point mutations at residues 197 
and 217 (although this residue is not present in the epitope, its
mutation will affect the epitope due
to physical proximity). The calculated value $p_{\rm epitope} = 2/6 = 0.33$.
Epitope C (residues 328--332, 334, 336, 338, 339, 341--344, 346, 347,
357--359, 366--370) presents one point mutation
at residue 370. The calculated value $p_{\rm epitope} = 1/23 = 0.043$.
We conclude that likely
epitope B is dominant, and epitope A is subdominant
for the A/Panama/2007/99 NA  protein, whereas epitope C is cryptic.

\section{Discussion}

In Figure \ref{fig1} are shown the predicted immune responses to the 
A/Fujian/411/2002
strain for the  hemagglutinin (green) and neuraminidase (red) dominant
epitopes after vaccination to A/Panama/2007/99. The predicted values 
for hemagglutinin
lie in the region of moderate immune response,
and so consistent with the WHO 
data \cite{WHO} one would expect some degree of cross-strain
protection. However, the predicted 
immune response to the dominant neuraminidase epitope
is in the region of  original antigenic sin.
In the design of the 2003--2004 influenza vaccine,
neither the cross activity nor the immune response were measured for
the A/Fujian/411/2002 NA protein in response to A/Panama/2007/99 NA
vaccination \cite{WHO}.  Upon analysis of actual effectiveness
of the 2003--2004 vaccine,
there does appear to have been a modest degree of original
antigenic sin, or negative vaccine effectiveness \cite{CDC}.

Summarizing our findings, it would appear that 
for the hemagglutinin protein of A/Panama/2007/99,
epitope B is dominant, and vaccination
gives modest protection to the A/Fujian/411/2002 strain.
For the neuraminidase protein of A/Panama/2007/99,
it would appear that epitope B is also dominant, and vaccination 
may increase the susceptibility to the A/Fujian/411/2002 strain.
Taken in aggregate, these results, Figure \ref{fig1}, 
suggest that the 2003--2004 flu vaccine would have 
essentially no effect against the circulating
A/Fujian/411/2002 strain, in agreement with clinical findings \cite{CDC}
and in disagreement with early expectations \cite{WHO}.

In conclusion, we suggest that strains related by antigenic drift 
be compared by measuring differences in the epitope 
regions of the hemagglutinin and neuraminidase
proteins and not by differences in the whole sequence or 
phylogeny as is presently done. This particular point is supported by the 
correlations of seasonal flu severity with epitope 
antigenic drift shown 
in Figure \ref{fig2}. Thus, there is a need for a detailed 
characterization of the epitope regions in the different 
influenza strains. In particular, a precise determination of
which epitopes are dominant in the proposed vaccine strain
and an experimental measure of $p_{\rm epitope}$ and of cross activity
between the proposed vaccine strain and the circulating strains
would be highly productive. In absence of this determination, we suggest to 
estimate the dominant epitope as that which shows the most antigenic drift 
\cite{Fitch,Bush,Plotkin,Fitch2}, as we do in the present work.  
From either epitope sequence drift or cross activity, 
Figure \ref{fig1} can be used
to estimate the degree of the immune response, which is 
a non-linear and non-monotonic function of the measured data.
We believe that this quantitative epitope analysis 
should be incorporated as part of the regular 
protocol for construction of the annual influenza vaccine.

\section*{Acknowledgments}

We thank Nancy Cox, Director of the WHO Collaborating Center for 
Surveillance, Epidemiology and Control of Influenza, 
Centers for Disease Control and Prevention (Atlanta),
for helpful
discussions. We also thank Xijan Xu 
from the Strain Surveillance Section, Influenza Branch G16,
Centers for Disease Control and Prevention (Atlanta), for providing the 
A/Fujian/411/2002 neuraminidase sequence.
This research was supported by the U.\ S.\ National Institutes of Health.

ETM carried out the sequence and structural analysis.
MWD conceived of the study and participated in its
design and coordination.  All authors read and approved the final
manuscript.

Correspondence and requests for materials should be addressed to 
mwdeem@rice.edu.

\bibliography{influenza}

\begin{thebibliography}{10}
\expandafter\ifx\csname url\endcsname\relax
  \def\url#1{\texttt{#1}}\fi
\expandafter\ifx\csname urlprefix\endcsname\relax\def\urlprefix{URL }\fi

\bibitem{CDC}
S.~Dolan, A.~C. Nyquist, D.~Ondrejka, J.~Todd, K.~Gershman, J.~Alexander,
  C.~Bridges, J.~Copeland, F.~David, G.~Euler, P.~Gargiullo, K.~Kenyan,
  Z.~Moore, J.~Seward, N.~Jain, Preliminary assessment of the effectiveness of
  the 2003--4 inactivated influenza vaccine - {C}olorado, {D}enver 2003,
  Centers for Disease Control and Prevention Morbidity and Mortality Weekly
  Report 53~(1) (2004) 8--11.

\bibitem{WHO}
N.~Cox, A.~Balish, L.~Brammer, K.~Fukuda, H.~Hall, A.~Klimov, S.~Lindstrom,
  J.~Mabry, G.~Perez-Oronoz, A.~Postema, M.~Shaw, C.~Smith, K.~Subbarao,
  T.~Wallis, X.~Xijan, Information for the vaccines and related biological
  products advisory committee, {CBER}, {FDA}, 2003, {WHO} Collaborating Center
  for Surveillance Epidemiology and Control of Influenza.

\bibitem{Deem}
M.~W. Deem, H.-Y. Lee, Sequence space localization in the immune system
  response to vaccination and disease, Phys. Rev. Lett. 91 (2003) 068101.

\bibitem{Epitope2}
G.~M. Air, M.~C. Els, L.~E. Brown, W.~G. Laver, R.~G. Webster, Location of
  antigenic sites on the three-dimensional structure of the influenza {N}2
  virus neuraminidase, Virology 145 (1985) 237--248.

\bibitem{MMWR91}
{Centers for Disease Control}, Update: Influenza activity --- {U}nited {S}tates
  and worldwide, 1991--92 season, and composition of the 1992--93 influenza
  vaccine, Centers for Disease Control and Prevention Morbidity and Mortality
  Weekly Report 41~(18) (1992) 315--317,323.

\bibitem{MMWR92}
{Centers for Disease Control}, Current trends update: Influenza activity ---
  {U}nited {S}tates and worldwide, 1992--93 season, and composition of the
  1993--94 influenza vaccine, Centers for Disease Control and Prevention
  Morbidity and Mortality Weekly Report 42~(10) (1993) 177--180.

\bibitem{MMWR93}
{Centers for Disease Control}, Current trends update: Influenza activity ---
  {U}nited {S}tates and worldwide, 1993--94 season, and composition of the
  1993--94 influenza vaccine, Centers for Disease Control and Prevention
  Morbidity and Mortality Weekly Report 43~(10) (1994) 179--183.

\bibitem{MMWR94}
{Centers for Disease Control}, Update: Influenza activity --- {U}nited {S}tates
  and worldwide, 1994--95 season, and composition of the 1995--96 influenza
  vaccine, Centers for Disease Control and Prevention Morbidity and Mortality
  Weekly Report 44~(15) (1995) 292--295.

\bibitem{MMWR95}
{Centers for Disease Control}, Update: Influenza activity --- {U}nited {S}ates
  and worldwide, 1995--96 season, and composition of the 1996--97 influenza
  vaccine, Centers for Disease Control and Prevention Morbidity and Mortality
  Weekly Report 45~(16) (1996) 326--329.

\bibitem{MMWR96}
{Centers for Disease Control}, Update: Influenza activity --- {U}nited {S}tates
  and worldwide, 1996--97 season, and composition of the 1997--98 influenza
  vaccine, Centers for Disease Control and Prevention Morbidity and Mortality
  Weekly Report 46~(15) (1997) 325--330.

\bibitem{MMWR97}
{Centers for Disease Control}, Update: Influenza activity --- {U}nited {S}tates
  and worldwide, 1997--98 season, and composition of the 1998--99 influenza
  vaccine, Centers for Disease Control and Prevention Morbidity and Mortality
  Weekly Report 47~(14) (1998) 280--284.

\bibitem{MMWR98}
{Centers for Disease Control}, Update: Influenza activity --- {U}nited {S}tates
  and worldwide, 1998--99 season, and composition of the 1998--99 influenza
  vaccine, Centers for Disease Control and Prevention Morbidity and Mortality
  Weekly Report 48~(18) (1999) 374--378.

\bibitem{MMWR99}
{Centers for Disease Control}, Update: Influenza activity --- {U}nited {S}tates
  and worldwide, 1999--2000 season, and composition of the 2000--01 influenza
  vaccine, Centers for Disease Control and Prevention Morbidity and Mortality
  Weekly Report 49~(17) (2000) 375--381.

\bibitem{Lanlha2}
C.~Macken, H.~Lu, J.~Goodman, L.~Boykin, The value of a database in
  surveillance and vaccine selection, in: A.~D. M.~E. Osterhaus, N.~Cox, A.~W.
  Hampson (Eds.), Options for the Control of Influenza {IV}, Elsevier Science,
  2001, accession number {ISDN38157}. http://www.flu.lanl.gov/.

\bibitem{Fazekas1}
S.~{Fazekas de St. Groth}, R.~G. Webster, Disquisition on original antigenic
  sin: Evidence in man, J. Exp. Med. 124 (1966) 331--345.

\bibitem{Davenport}
F.~M. Davenport, A.~V. Hennessy, T.~Francis, Epidemiologic and immunologic
  significance of age distribution of antibody to antigenic variants of
  influenza virus, J. Exp. Med. 98 (1953) 641--656.

\bibitem{Bellini}
W.~J. Bellini, A.~P. Rota, Genetic diversity of wild-type measles viruses:
  {I}mplications for global measles elimination programs, Emerging Infectious
  Diseases 4 (1998) 29--35.

\bibitem{Lanlha1}
Accession number {ISDNCDA001} from the Influenza Sequence
  Database\cite{Lanlha2}.

\bibitem{Lanlha}
Hemagglutinin H3 epitope structural mapping from the Influenza Sequence
  Database\cite{Lanlha2}.

\bibitem{Bush99}
R.~M. Bush, W.~M. Fitch, C.~A. Bender, N.~J. Cox, Positive selection on the
  {H}3 hemagglutinin gene of human influenza virus {A}, Mol. Biol. Evol. 16
  (1999) 1457--1465.

\bibitem{Cox}
N.~J. Cox, C.~A. Bender, The molecular epidemiology of influenza viruses,
  Semin. Virol. 6 (1995) 359--370.

\bibitem{Fitch}
W.~M. Fitch, J.~M. Leiter, X.~Li, P.~Palese, Positive {D}arwinian evolution in
  human influenza {A} viruses, Proc. Natl. Acad. Sci. USA 88 (1991) 4270--4274.

\bibitem{Bush}
R.~M. Bush, C.~A. Bender, K.~Subbarao, N.~J. Cox, W.~M. Fitch, Predicting the
  evolution of human influenza {A}, Science 286 (1999) 1921--1925.

\bibitem{Plotkin}
J.~B. Plotkin, J.~Dushoff, Codon bias and frequency-dependent selection on the
  hemagglutinin epitopes of influenza {A} virus, Proc. Natl. Acad. Sci. USA 100
  (2003) 7152--7157.

\bibitem{Fitch2}
W.~M. Fitch, R.~M. Butch, C.~A. Bender, K.~Subbarao, N.~J. Cox, Predicting the
  evolution of human influenza {A}, The Journal of Heredity 91 (2000) 183--185.

\bibitem{Kostolansky}
F.~Kostolansk\'{y}, E.~Vare\v{c}kova, T.~Bet\'{a}kov\'{a}, V.~Mucha, G.~Russ,
  S.~A. Wharton, The strong positive correlation between effective affinity and
  infectivity neutralization of highly cross-reactive monoclonal antibody
  {IIB}4, which recognizes antigenic site {B} on influenza {A} virus
  haemagglutinin, Journal of General Virology 81 (2000) 1727--1735.

\bibitem{Lanlna2}
Personnal communication, Xu Xijan.

\bibitem{Lanlna1}
Accession number {AJ457937} from the Influenza Sequence Database\cite{Lanlha2}.

\bibitem{Tulip}
W.~R. Tulip, J.~N. Varghese, A.~T. Baker, A.~van Donkelaar, W.~G. Laver, R.~G.
  Webster, P.~M. Colman, Refined atomic structures of {N}9 subtype influenza
  virus neuraminidase and escape mutants, J. Mol. Bio. 221 (1992) 487--497,
  {PDB} accession number 3NN9.

\bibitem{Air1995}
J.~J. Jedrzejas, S.~Singh, W.~J. Brouillette, W.~G. Laver, G.~M. Air, M.~Luo,
  {S}tructures of aromatic inhibitors of influenza {A} virus neuraminidase,
  Biochemistry 34 (1995) 3144--3151, {PDB} accession number {1IVD}.

\bibitem{Gulati}
U.~Gulati, C.-C. Hwang, L.~Venkatramani, S.~Gulati, S.~Stray, J.~T. Lee, W.~G.
  Laver, A.~Bochkarev, A.~Zlotnick, G.~Air, Antibody epitopes on the
  neuraminidase of a recent {H}3{N}2 influenza virus {(A/Memphis/31/98)},
  Journal of Virology 76 (2002) 12274--12280.

\bibitem{Smith}
D.~J. Smith, S.~Forrest, D.~H. Ackley, A.~S. Perelson, Variable efficacy of
  repeated annual influenza vaccination, Proc. Natl. Acad. Sci. USA 96 (1999)
  14001--14006.

\end{thebibliography}

\clearpage

\begin{center}
{\bf Figures}
\end{center}

\begin{figure}[h]
\caption{
The evolved affinity constant to a second antigen after exposure to
an original antigen whose epitope region
differs by probability $p_{\rm epitope}$ (solid line).
The dashed line represents the affinity constant without previous exposure.
In green are shown the responses at the values of the
differences
between the A/Panama/2007/99 (vaccine) and A/Fujian/411/2002
(circulating) strains for the B and E
hemagglutinin epitopes.
In red are shown the responses at the values of the
difference between the A/Panama/2007/99 (vaccine) and A/Fujian/411/2002
(circulating) strains for the A and B
neuraminidase epitopes.
Dominant epitopes are shown in bold.
The clinical outcome is an average of the response to the
HA and NA proteins, and in purple is shown the immune response at
the average difference for the dominant HA and NA epitopes (see text).
The effectiveness of the 2003--2004 flu vaccine was marginal at best.
In inset is shown the cross affinity of
the memory sequences for the mutated antigen, often measured
biochemically and distinct from the evolved immune response.
As if often found, the cross activity decreases exponentially
with antigenic drift \cite{Smith}.
\label{fig1}
}
\end{figure}
\bigskip
\bigskip

\begin{figure}[h]
\caption{
a) Correlation between influenza seasonal severity (see text)
and hemagglutinin antigenic drift, calculated by epitope 
analysis. Least-squares regression analysis yields the linear fit 
$y = 1.5425 + 5.9162  \, p_{\rm epitope}$, with a correlation coefficient
$R = 0.54432$.  
b) Correlation between influenza seasonal severity (see text)
and hemagglutinin antigenic drift, calculated by whole 
sequence analysis. Least-squares regression analysis yields the linear fit
$y = 2.0041 + 11.684  \, p_{\rm sequence}$, with a correlation coefficient
$R = 0.2183$.
\label{fig2}
}
\end{figure}
\bigskip
\bigskip

\begin{figure}[h]
\caption{
a) Shown are the dominant B (top) and subdominant E (middle) epitope of the
hemagglutinin protein in space-filling format \cite{Lanlha}.
b) Shown are the dominant B (right) and subdominant A (left) epitope of the
neuraminidase protein in space-filling format \cite{Air1995}.
The rest of the proteins are shown in ribbon format.
\label{fig3}
}
\end{figure}

\vfill

\clearpage

\centering
\leavevmode
\psfig{file=fig1.eps,height=4in,clip=}

\vspace*{0.5in}Figure~\ref{fig1}.
Munoz and Deem, ``Epitope analysis \ldots."

\clearpage

\centering
\leavevmode
\psfig{file=fig2a.eps,height=3in,clip=}

\centering
\leavevmode
\psfig{file=fig2b.eps,height=3in,clip=}

\vspace*{0.5in}Figure~\ref{fig2}.
Munoz and Deem, ``Epitope analysis \ldots."

\clearpage

\centering
\leavevmode
{\Large \sf a) }
\psfig{file=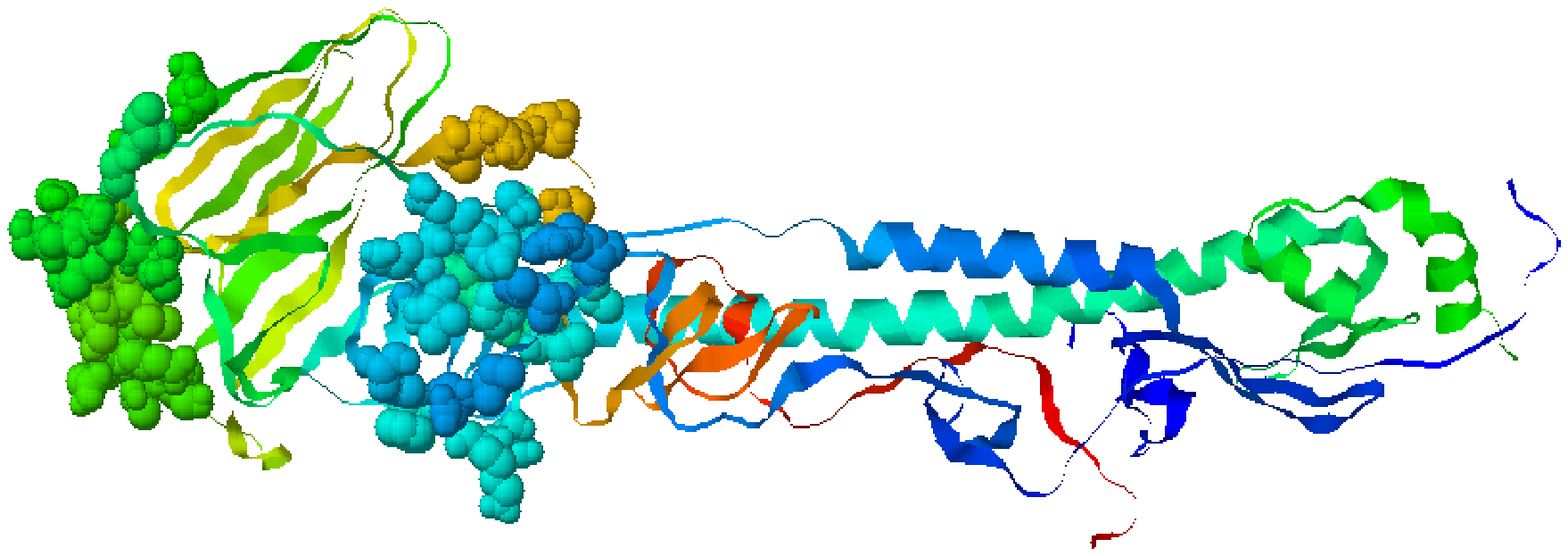,width=2in,clip=,angle=-90}
{\Large \sf b) }
\psfig{file=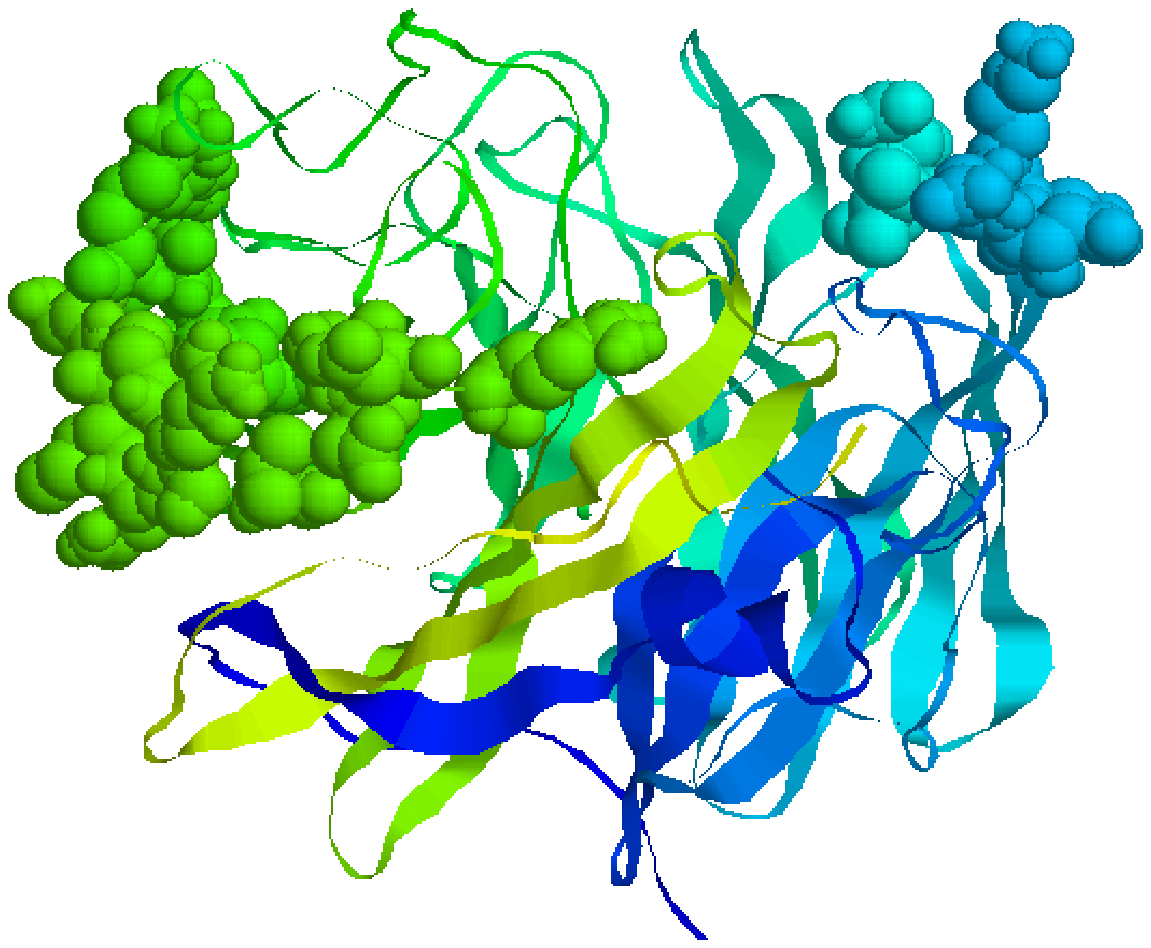,width=2in,clip=,angle=-90}

\vspace*{0.5in}Figure~\ref{fig3}.
Munoz and Deem, ``Epitope analysis \ldots."

\end{document}